# A Review of Platforms for the Development of Agent Systems

Constantin-Valentin Pal[1], Florin Leon[1], Marcin Paprzycki[2], Maria Ganzha[2]

[1] *Faculty of Automatic Control and Computer Engineering*
"Gheorghe Asachi" Technical University of Iaşi, Romania

[2] *Systems Research Institute*
*Polish Academy of Sciences*

valentin.pal@tuiasi.ro, florin.leon@tuiasi.ro,
paprzyck@ibspan.waw.pl, maria.ganzha@ibspan.waw.pl

**Abstract**
Agent-based computing is an active field of research with the goal of building autonomous software of hardware entities. This task is often facilitated by the use of dedicated, specialized frameworks. For almost thirty years, many such agent platforms have been developed. Meanwhile, some of them have been abandoned, others continue their development and new platforms are released. This paper presents a up-to-date review of the existing agent platforms and also a historical perspective of this domain. It aims to serve as a reference point for people interested in developing agent systems. This work details the main characteristics of the included agent platforms, together with links to specific projects where they have been used. It distinguishes between the active platforms and those no longer under development or with unclear status. It also classifies the agent platforms as general purpose ones, free or commercial, and specialized ones, which can be used for particular types of applications.

**Keywords:** agent systems, agent-based systems, multiagent systems, agent platforms, modeling, simulation, swarm intelligence, artificial intelligence

## 1. Introduction

It is already almost 10 years since the last comprehensive overview of tools for the development of agent systems has been published (Bădică et al., 2011). Since agent systems remain an active research area, and software agents start to be applied in new domains (e.g., Savaglio et al., 2020) we believe that it is time to reflect on the current state of the art in platforms for development of agent systems. Let us start our work with highlighting key developments in the area of agent systems.

### 1.1. A Short History of Agent Systems

Perhaps, one of key moments in development agent systems was the Workshop on Distributed Artificial Intelligence, held at MIT in June 1980, where 22 people presented their research results and ideas (Davis, 1980). Next, 1980s saw the first attempts at defining the main concepts of the agents domain and, among them, one can recognize key issues that are still studied today.

In 1984, Axelrod showed how cooperation can emerge from the interaction of selfish entities, without centralized control (Axelrod, 1984). He discussed several strategies in the iterated prisoner's dilemma, a game theoretical problem which remains to be a topic of interest in political and social studies or evolutionary biology.

In 1985, the actor model was proposed by Agha and Hewitt (1985). With all conceptual limitations, actors can be seen as a "simpler version" of agents. An actor is reactive, as it only responds to messages it receives, but it can nevertheless be used to simulate more advanced behavior, such as perceiving the environment and initiating new actions when its internal state changes in some way.

In 1986, Brooks stated his rejection of the logical, symbolic approach to intelligence, prevalent at that time, and proposed the subsumption architecture (Brooks, 1986), implemented in four robots capable of seemingly intelligent behavior, without any symbolic internal representation of the environment.

At the beginning of the 1990s, Maes worked together with Brooks at MIT, to design robots for building a base on the Moon (Brooks et al., 1990), and then she turned to software agents for personalized information filtering (Sheth & Maes, 1993), later abstracted into what we commonly view now as agents – an autonomous entity (Maes, 1993; Maes et al., 1995).

Also in 1986, the first (micro-)simulations were conceived, namely in the pursuit domain, where a number of predators aimed at encircling, or capturing, a prey (Benda, Jagannathan & Dodhiawala, 1986). Note that simulations in different domains are presently one of the more popular applications of multiagent systems. Even if we can consider proper simulation only in a computer-based settings, interest in this area was even older, e.g. the paper-based analysis of segregation in the Schelling's community (Schelling, 1969), in a grid-based environment reminiscent of even older cellular automata, invented by von Neumann and Ulam in 1940, and made famous by Conway's "Game of Life" proposed in 1970.

Based on the psychological studies of practical reasoning (Bratman, 1987), Georgeff and Lansky developed in 1987 a so-called "Procedural Reasoning System" (PRS), based on the Belief-Desire-Intention (BDI) model for intelligent agents (Georgeff & Lansky, 1987). A formalization of the BDI architecture, which is one of the most popular in the agent domain, can be found in (Rao & Georgeff, 1995). The model explicitly includes the agent's beliefs about the state of the environment and its own. BDI incorporates the concepts of establishing the means available to reach a certain goal and creating a plan whose actions can be performed sequentially.

While the first agents were built using the available means, e.g. in Lisp, as the complexity of applications increased, and experience was gathered, the need of specialized frameworks for constructing agents was recognized. One of the first attempts was AGENT0 (Torrance & Viola, 1991), a language that incorporated the idea of agent-oriented programming (Shoham, 1993), where the agent has full control over its own state (beliefs, capabilities) and behavior (responding to messages, commitments).

The 1990s witnessed a steady stream of development in the agent systems domain. Here, we can identify separate areas of interest. There were advances in the theoretical study of such systems. The ARCHON project (Jennings, 1994) proposed a general-purpose architecture, which could be used to facilitate cooperative problem-solving in industrial applications. KQML (Knowledge Query and Manipulation Language) was put forward, within the DARPA Knowledge Sharing Effort (Finin et al., 1994), and proposed a standardized way for agent communication which separated the intent of the message in the form of a so-called "performative" (inspired by the speech act theory (Austin, 1962)) from the actual content of the message. This effort was further refined by FIPA ACL, Agent Communication Language (FIPA, 1997). Software agents were described and discussed in several influential papers by Wooldridge and Jennings (Jennings & Wooldridge, 1996; Wooldridge, 1996; Wooldridge, 1997). A specific agent-based language, AgentSpeak, inspired by PRS, was formally introduced (d'Inverno & Luck, 1998).

Here, interest emerged in autonomous cars, seen as autonomous agents. Agent architectures were proposed, e.g. Touring Machine (Ferguson, 1992), InteRRaP (Müller & Pischel, 1993), and the first self-driving vehicles were put to the test in competitions, e.g. ALVINN (Pomerleau, 1995) and later Nomad rover (Jordan, Andreas & Makshtas, 2001) or Stanley (Thrun et al., 2007).

Another class of applications for multiagent systems is related to social simulations, usefulness of which continues. In this respect, we can mention the Sugarscape emergent economic model (Epstein & Axtell, 1996), the evolution of social corruption (Hammond, 2000), the study of the population dynamics of Cyber-Anasazi (Axtell et al., 2002) and a model for civil violence, leading to "artificial genocide" (Epstein, 2002).

Reflecting on the connection to the industry, we can mention the AgentLink European project (three phases, 1998-2003), which studied the application of agent systems in domains such as telecommunications, information management, electronic commerce or manufacturing. COST Agreement Technologies We can see that the vision of software agents from the AgentLink is materializing now, in the world of the Internet of Things, IoT (Savaglio et al., 2020).

Another European initiative was the COST Agreement Technologies action, which aimed at coordinating efforts on a new paradigm for next generation distributed systems, based on the concept of agreement between computational agents, i.e. computer systems in which autonomous software agents negotiate with one another, typically on behalf of humans, in order to come to mutually acceptable agreements.

After a period of slow decline at the beginning of the new millennium (although the research community has remained active throughout), we can presently see a rebirth of interest, with new understanding of the importance of agent systems, especially in domains such as autonomous cars and drones, including those aimed at delivering goods to customers, simulations (e.g. evacuation behavior in emergency situations (Teo et al., 2015)) or smart cities and IoT, where new technology advances in the communication infrastructure (e.g. 5G) make further automation and interconnectedness of intelligent devices possible.

**1.2. Applications of Agent Systems**

This brief history of the field leads us to identify two main approaches in the development of agent systems. On one hand, we can consider agents as a metaphor and implement agent systems using any programming environment, usually an object-oriented one. Still, this requires work that is not really needed, in dealing with the specific aspects of agent systems. On the other hand, a better way seems to be the use of a dedicated agent platform. Agent-based modeling and simulation software of this kind take away many of the complexities of the modeling and simulation implementation. They allow the user to focus on studying the phenomena that arise from the interaction of agents, e.g. emergent behaviors, or on addressing problems that are difficult to solve in a centralized manner.

Among the main classes of applications where agent systems have been and are used, we can mention the following:

- *Social simulations:* various scenarios, such as those presented in the previous section;
- *Mobility simulations:* traffic situations, such as the avoidance of traffic jams, light control, route choice, e.g. (Czura et al., 2014), ground transportation, mobility planning systems, urban planning based on accessibility studies with dynamic populations e.g. (Tian & Qiao, 2014), microscopic pedestrian crowds, e.g. (Wang et al., 2015), or mapping passenger flow for market improvement and evacuation of buildings, flight or air-traffic control in aviation, e.g. (Bongiorno et al., 2013; Horio et al., 2015) etc.;
- *Physical entities:* robots or self-driving vehicles (cars, drones) seen as agents;
- *Environment and ecosystems:* simulations in ecology, e.g. (Ayllón et al., 2016), biology, climate models, human and nature interaction (sometimes using geographic information systems), epidemiology (the spread of infections or disease), e.g. (García et al., 2017);
- *Organizational simulations:* planning and scheduling, enterprise and organizational behavior, workflow simulations, e.g. (Prenkert & Følgesvold, 2014);
- *Economic studies:* business, marketing, economics (e.g. price forecasting in real world markets);
- *Medical applications:* personalized healthcare or hospital management, e.g. (Sulis & Di Leva, 2017);
- *Industrial simulations:* manufacturing and production, including with the use of holons, e.g. (Parv et al., 2019);
- *Military applications:* military-combat simulations, air-defense scenarios, e.g. (Lee et al., 2018).

On the other hand, there are agent-based applications for distributed computing, e.g. in cloud computing, virtualized data centers, large-scale parallel or distributed computing clusters and high performance supercomputers (Kiourt & Kalles, 2015; Taylor et al., 2018).

A recent book that focuses on the practical aspects related to agent systems is (Ganzha & Lakhmi, 2013).

The types of applications mentioned above are displayed more concisely in Figure 1.

Finally, we can also mention the application of agents in games or movie making, games or graphic engines, e.g. Massive (Massive Software, 2017), Unity-ML (Booth & Booth, 2019; Unity Technologies, 2020) or DeepMotion (DeepMotion, 2020).

## 1.3. The Structure of the Review

Taking into account the breadth and depth of applications of software agents, as well as the continuous development and growing maturity of agent platforms, not to mention recent creation of new ones, we have decided that the time has come to deliver an up-to-date overview of existing agent platforms, combined with the summary of more recently departed ones.

With the decrease of interest after 2005, the majority of these platforms have been more or less abandoned. This is why it is important to reflect on what is available today – and this is the goal of our present work.

We would like to stress that this is not an introduction in the field of multiagent systems. The readers interested in an more introductory perspective are invited to study some of the general books on the subject, such as (Wooldridge, 2009; Weiss, 2013).

The current paper presents a list of more up-to-date software, free or commercial, the latter sometimes with special discounts or free offerings for academia. In addition, this survey also includes some platforms oriented more towards the Artificial Intelligence domain. These platforms can be used for multiagent problem solving, or for studying the emergent behavior of intelligent multiagent systems in various domains like arcade games, robotics, autonomous vehicles, or agents that are capable of reinforcement learning. The structure of the paper is graphically presented in Figure 2.

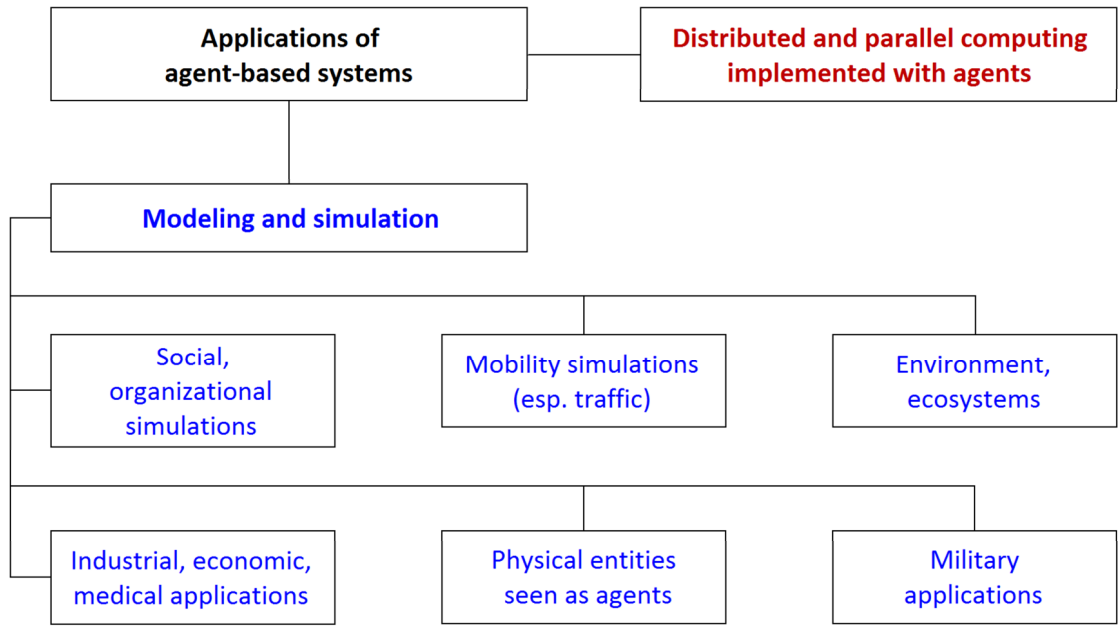

*Figure 1. Types of applications of agent systems*

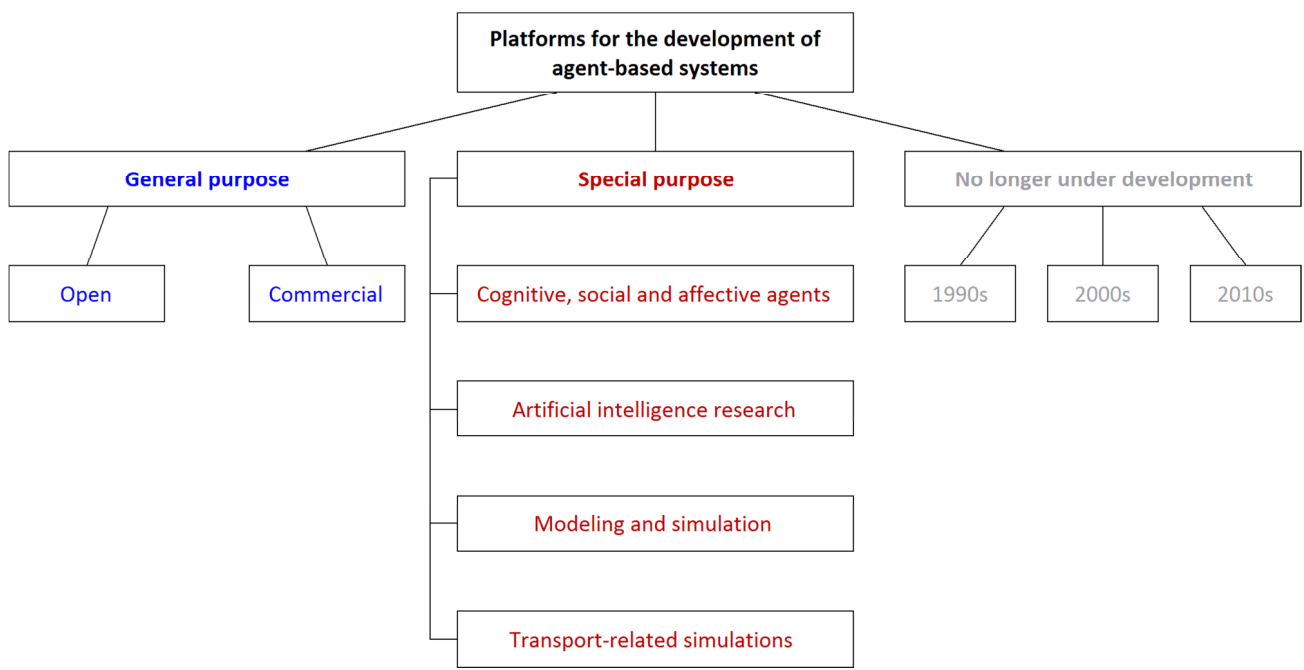

*Figure 2. The main structure of the current review*

## 2. Related Work: Earlier Reviews of Agent Platforms

As the field of agent systems matured, a number of authors have reviewed the landscape of this type of software. Some reviews are confined within a specific application domain, like electricity market (Zhou, Chan & Chow, 2007), marketing (Negahban & Yilmaz, 2014), networking (Niazi & Hussain, 2009), land use (Groeneveld et al., 2017). (Dorri et al., 2018) discusses aspects like definitions, applications, features, challenges, communications, evaluation, but does not directly present a comprehensive list of software platforms that can be used to implement agent systems. Older reviews of software in this space can be found in (Railsback, Lytinen & Jackson, 2006; Bordini et al., 2006). Allan (2010) reviews the development at the time of the paper and, beside software platforms, discusses the domains of application with use cases from physics, chemistry, biology, cyber-security, social modeling, economics, environment. Newer papers, e.g. (Kravari & Bassiliades, 2015; Abar et al., 2017) also review a wide range of software tools, the various features of the tools and types of licenses. Leon, Paprzycki and Ganzha (2015) presents the list of the most important software at the time of the writing and some methodologies for agent system development.

## 3. General Purpose Platforms

Let us start with general purpose agent platforms. We summarize them in the following two tables. In each table, we provide links for the website of the platform and its applications in research projects. We consider that, in this way, the links are more easy to use, compared to the situation where they were included as references at the end of the paper. Following these two tables, we summarize specific features of mentioned platforms. In this case, we may mention some references of publications that presented, or described, applications developed using a particular framework.

The platforms from this section can be used in a variety of domains, since they are not "domain focused". Various programming languages are used for the implementation of the platforms, and there are also some platforms that can be leveraged just by manually interacting with a graphical user interface. Some platforms offer modular and hierarchical modeling, based on reusable components, and they range from low to very high scale simulation with millions of agents.

### 3.1. Open-Source, Free Software

Table 1 contains a list with the general purpose platforms that seem to be the most popular open-source or free license software.

*Table 1. General purpose platforms - open-source, free software*

| No. | Name | Programming language | Website. Projects and applications | License | Description |
|---|---|---|---|---|---|
| 1 | **ActressMas** | C# | http://florinleon.byethost24.com/actressmas<br>*Applications (several multiagent algorithms and protocols implemented):*<br>https://github.com/florinleon/ActressMas | Open source | Used for teaching multiagent protocols and algorithms. offers implementations of various popular multiagent protocols and algorithms |
| 2 | **Agents.jl** | Julia | https://github.com/JuliaDynamics/Agents.jlhttps://doi.org/10.21105/joss.01611 | Open source | General purpose, grid-based environments; 1D, 2D, 3D; distributed simulations |
| 3 | **AgentScript** | Javascript | https://github.com/backspaces/agentscript | Open source | General purpose, based on NetLogo semantics |
| 4 | **CoSMoSim** | GUI based | https://sourceforge.net/projects/cosmosim/<br>https://acims.asu.edu/software/cosmos/ | Open source (Java) | Component-based, modular, hierarchical modeling; DEVS, cellular automata and XML models |
| 5 | **DEVS-Suite** | GUI based | https://acims.asu.edu/software/devs-suite/<br>https://sourceforge.net/projects/devs-suitesim/ | Open source (Java) | Rich visual modeling, component-based and cellular automata simulator, hierarchical models, superdense time data trajectories |
| 6 | **Evoplex** | C++ | https://evoplex.org/en/ | Open source | An agent is represented as a node in a network; Evolutionary Graph Theory, Evolutionary Dynamics, Game Theory, Cellular Automata, Complex Adaptive Systems. |
| 7 | **Gama** | Java GAML | https://github.com/gama-platform<br>http://gama-platform.org<br>https://code.google.com/p/gama-platform/<br>*Projects:*<br>https://gama-platform.github.io/wiki/Projects | Open source (Java) | Complete modeling and simulation development environment for building spatially explicit multiagent simulations |
| 8 | **FLAME** | C/C++ | http://flame.ac.uk/<br>https://github.com/FLAME-HPC/<br>http://flame.ac.uk/projects/ | Open source | It generates a complete agent-based application which can be compiled and built on most computing systems ranging from laptops to HPC supercomputers. |
| 9 | **FLAME GPU** | C for CUDA, C-based scripting | http://www.flamegpu.com/home | Open source (C/C++) | Graphics Processing Unit (GPU) extension to the FLAME framework |
| 10 | **Insight Maker** | Runs in the browser and modeling is done through the browser UI | https://insightmaker.com/ <br> *Projects:* <br> https://insightmaker.com/new | Open source (Qt) | System dynamics, agent-based modeling in the browser. |
| 11 | **JaCaMo** | AgentSpeak (Jason) | http://jacamo.sourceforge.net/ <br> http://cartago.sourceforge.net/ | Open source | Autonomous agents, environment artifacts, multiagent organizations |
| 12 | **JADE** | Java, C# (JADE LEAP) | http://jade.tilab.com/ | Open source (Java) | FIPA-compliant middleware, graphical debugging and deployment tools |
| 13 | **JADEX** | Java | https://www.activecomponents.org/#/download | Open source (Java) | Rational agents on top of JADE, BDI |
| 14 | **Janus, SARL** | SARL, interoperable with Java | http://www.janusproject.io/ <br> https://github.com/janus-project <br> http://www.sarl.io/ | Open source (Java) | Agent-oriented SARL language, fundamental abstractions for dealing with concurrency, distribution, interaction, decentralization, reactivity, autonomy and dynamic reconfiguration |
| 15 | **JAS-mine** | Java | http://www.jas-mine.net/ | Open source | Discrete-event simulation, including agent-based and micro-simulation models. Integration with RDBMS (relational database management tools |
| 16 | **MADKIT** | Java | http://www.madkit.org/ | Open source | AGR (Agent/Group/Role) organizational model: agents play roles in groups and thus create artificial societies. |
| 17 | **MASON** | Java | https://cs.gmu.edu/~eclab/projects/mason/ <br> *Projects:* <br> https://github.com/eclab/mason/ <br> https://cs.gmu.edu/~eclab/projects/mason/#Projects <br> *Manual:* <br> https://cs.gmu.edu/~eclab/projects/mason/manual.pdf | Open source | Discrete event multiagent simulation; 2D and 3D visualization |
| 18 | **MASS** | Java, C++, Cuda | http://depts.washington.edu/dslab/MASS/ | Open source (Java, C++) | Parallel-computing library for multiagent and spatial simulation over a cluster of computing nodes. |
| 19 | **Mesa** | Python 3+, recent code, | https://mesa.readthedocs.io/en/master/overview.html <br><br> https://github.com/projectmesa/mesa <br><br> https://www.researchgate.net/publication/328774079_Mesa_An_Agent-Based_Modeling_Framework | Open source, Apache2 licensed (Python) | Python 3 alternative to NetLogo, Repast, MASON. |
| 20 | **MOOSE** | C++ | https://www.mooseframework.org/ <br> *Code:* <br> https://github.com/idaholab/moose | Open source (C++) | High-scale Multiphysics object-oriented simulation environment. |
| 21 | **Orleans** | C# | http://research.microsoft.com/en-us/projects/orleans/<br>https://github.com/dotnet/orleans<br>*Internal Microsoft projects:*<br>http://research.microsoft.com/en-us/projects/orleans/ | Open source | Distributed high-scale computing applications, without the need to learn and apply complex concurrency or other scaling patterns. |
| 22 | **Repast** | Java, Python, C#/.NET, C++, ReLogo, Groovy | https://repast.github.io/<br>*Projects:*<br>http://www2.econ.iastate.edu/tesfatsi/repastsg.htm#Project<br>http://www2.econ.iastate.edu/tesfatsi/repastsg.htm#Project | Open source (Java, C++) | Agent-based modeling and simulation that can run on large computing clusters and supercomputers. |
| 23 | **SeSAm** | GUI programming | http://www.simsesam.de/ | LGPL license (Java) | Distributed agent-based modeling and simulation |
| 24 | **SLAPP** | Python | https://github.com/terna/SLAPP3 | Open source | Swarm-like agent protocol |
| 25 | **SPADE** | Python | https://pypi.python.org/pypi/SPADE | Open source | Multiagent and Organizations Platform based on the instant messaging XMPP/Jabber technology |
| 26 | **SpaDES** | R | http://spades.predictiveecology.org/<br>*Code:*<br>https://github.com/PredictiveEcology/SpaDES | GPL-3 | Spatially explicit discrete event simulation models |

**ActressMas** is an agent framework based on the Actor model, implemented by means of .NET asynchronous operations. Its main objective is simplicity and it is primarily dedicated to students learning about agent systems. It is one of the few agent frameworks available for C#. Many algorithms specific to agent systems were implemented, and also a traffic simulator that can be used to gather data for autonomous driving scenarios.

**Agents.jl** is a Julia framework for agent-based modeling (ABM). It provides a structure and components for implementing agent-based models, run them in batch, collect data, and visualize them. It provides default grids to run the simulations, including simple or toroidal 1D grids, simple or toroidal regular rectangular and triangular 2D grids, and simple or toroidal regular cubic 3D grids with von Neumann or Moore neighborhoods. The simulations can be run in parallel on multiple cores and their results can be explored interactively in "Data Voyager". Agents.jl is inspired by the Mesa framework for Python.

**AgentScript** is a minimalist agent-based modeling (ABM) framework based on NetLogo agent semantics. Its goal is to promote the agent-oriented programming model in a CoffeeScript/ JavaScript implementation.

**CoSMoSim** offers an integrated framework for model development, simulation and experimentation. Its unified logical, visual, and persistence framework supports specifying families of parallel cellular automata (CA), discrete event system simulation (DEVS), Statecharts, and XML-Schema models. As an integrated modeling and simulation environment, CoSMoS supports configuring input/output data monitoring and visualization for every model component. Logical specifications of models and their elements are stored in a Microsoft Access relational database. Code automatically generated from model components are systematically managed without any involvement by the user. While full Java code for the DEVS-Suite simulator is automatically generated for instances of coupled DEVS models, at present only partial code generation is supported for atomic parallel DEVS models. Once implementation of the atomic models are completed using a built-in editor, coupled models can be executed using the DEVS-Suite simulator.

**DEVS-Suite** is a Parallel DEVS Component-based and Cellular Automata simulator with support for: automating design of experiments in combination with generating superdense time data trajectories at run-time, hierarchical model libraries, animating models, synchronized run-time viewing for time-based trajectories and box-in-box hierarchical component and I/O messaging viewing. Capabilities for tracking, animation, playback, and area zooming are supported at scale. New concepts for component tracking (automated transducers) include mixed separate and stack time trajectories with naming, dynamic tracking log, and configuration interface. For Cellular Automata, independent tracking of any number of cells with independent start times is supported. The time-based trajectory plotting is enriched to handle zero-time advance at the start of simulation. This simulator engine package is strictly separated from model packages. The "Models" package is divided into the "CellularAutomata" and "Component" packages. User-defined model packages can be added alongside the provided model packages. The Component package is composed of a collection of packages, each providing a set of model components focusing on a class of general and specific system types including single-input-single-output, basic and multiprocessor architectures, and switch networks. The Cellular Automata package offers a collection of packages including the game of life, system biology chemotaxis, forest fire, and heat diffusion.

**Evoplex** is a fast, robust and extensible platform for developing agent-based models and agent systems on networks. Each agent is represented as a node and interacts with its neighbors, as defined by the network structure. Evoplex is a fast, multi-threaded, user-friendly, cross-platform and modular application. It was originally developed to tackle problems in the field of evolutionary computation and complex systems. However, it has been used in a wide range of scenarios, including: evolutionary graph theory, evolutionary dynamics, game theory, cellular automata and complex adaptive systems.

**GAMA** is a simulation platform, which aims at providing field experts, modelers, and computer scientists with a complete modeling and simulation development environment for building spatially explicit multiagent simulations. GAMA has been developed with a very general approach and can be used for many application domains. Some additional plugins had been developed to fit particular needs. Example of application domains where GAMA is mostly present are: transport, urban planning, epidemiology and environment. One can instantiate agents from any dataset, including GIS data, and execute large-scale simulations (up to millions of agents). Declare interfaces supporting deep inspections on agents, user-controlled action panels, multi-layer 2D/3D displays and agent aspects.

**FLAME** is a generic agent-based modeling system which can be used to develop applications in many areas. It generates a complete agent-based application which can be compiled and built on the majority of computing systems ranging from laptops to HPC supercomputers. Models are created based upon a model of computation called extended finite state machines. By defining agent-based models in this way the FLAME framework can automatically generate simulation programs that can run models efficiently on HPCs.

**FLAME GPU** is a high performance graphics processing unit (GPU) extension to the FLAME framework. It provides a mapping between a formal agent specifications with C-based scripting and optimized CUDA code. This includes a number of key ABM building blocks such as multiple agent types, agent communication and birth and death allocation. Agent-based modelers are able to focus on specifying agent behavior and run simulations without explicit understanding of CUDA programming or GPU optimization strategies. Massive agent populations can be visualized in real time as agent data is already located on the GPU hardware.

**Insight Maker** is a multi-method modeling solution packaged within a fluid and cohesive software environment. At one level, Insight Maker can be used purely to map out conceptual models: using causal loop diagrams or rich pictures to describe a system. In this mode, Insight Maker functions as a powerful diagramming tool that allows to illustrate a model and then easily share it with others. Once a model diagram has been created, behavior can be added to the different components using Insight Maker's simulation engine. Insight Maker supports two different modeling paradigms that together can describe most of the models imaginable. "System Dynamics" (sometimes called differential equation modeling or dynamical systems modeling) concerns itself with the high-level behavior of a system. It helps understanding the aggregate operations of a system on a macro-scale. It is good for cutting away unnecessary detail and focusing on what is truly important in a model. "Agent Based Modeling" allow the modeling of individual agents within a system. Where in "System Dynamics" the modeler looks only at the population, in "Agent Based Modeling" one can model each individual and explore the differences and interactions between these individuals.

**JaCaMo** is a framework for multiagent programming that combines three separate technologies, each of them being well-known on its own and developed for a number of years so they are fairly robust and fully-fledged: Jason, for programming autonomous agents, Cartago, for programming environment artifacts and Moise, for programming multiagent organizations. CartAgO (Common ARTifact infrastructure for AGents Open environments) is a framework for programming and executing virtual environments in multiagent programs. Jason is a platform for the development of agent systems that incorporates an agent-oriented programming (AOP) language.

**JADE** (Java Agent DEvelopment Framework) is a software framework implemented in Java. It simplifies the implementation of agent systems through a middleware that complies with the FIPA specifications and through a set of graphical tools that support the debugging and deployment phases.

**JADEX** (JADE extension) is a rational agent layer on top of JADE that allows for the easy development of rational agents. Intelligent agents follow a modeling paradigm, based on the notion of agents with mental states. The JADEX system realizes these concepts following the well-known belief, desire, intention (BDI) model at the design and implementation layer. The beliefs, goals and plans of the agents are defined in XML files, and the plan bodies are written in Java. It has a new version implemented only in Java called "ActiveComponents", which focuses on web services.

**Janus** is an open-source agent platform implemented in Java. It enables developers to create web, enterprise and desktop agent-based applications. It provides a set of features to develop, run, display and monitor agent-based applications. Janus-based applications can be distributed across a network. Janus can be used as an agent-oriented platform, an organizational platform, and/or a holonic platform. It natively manages the concept of recursive agents and holons. SARL is a general-purpose agent-oriented language. It aims at providing the fundamental abstractions for dealing with concurrency, distribution, interaction, decentralization, reactivity, autonomy and dynamic reconfiguration. Janus is a runtime environment for multiagent applications that supports the concepts of SARL.

**JAS-mine** is a Java platform that aims at providing a unique simulation tool for discrete-event simulations, including agent-based and micro-simulation models. With the aim to develop large-scale, data-driven models, the main architectural choice of JAS-mine is to use whenever possible standard, open-source tools already available in the software development community. The main added value of the platform are: integration of I/O communication tools, in the form of embedded relational database management systems (RDBMS) tools and automatic CSV table creation; advanced multi-run tools to facilitate the design of experiments; regression libraries that allow a complete separation of regression specifications from the code, and permit uncertainty analysis of the model outcome by bootstrapping the estimated coefficients across different simulation runs. JAS-mine allows the separation of data representation and management, which is automatically taken care of by the simulation engine, from the implementation of processes and behavioral algorithms, which should be the primary concern of the modeler. This results in quicker, more robust and more transparent model building.

**MADKIT** is a lightweight Java library for designing and simulating agent systems. MaDKit is designed to easily build distributed applications and simulations. In contrast to conventional approaches, which are mostly agent-centered, MaDKit follows an organization-centered approach without any predefined agent models. MaDKit is built upon the AGR (Agent/Group/Role) organizational model, i.e. agents play roles in groups and thus create artificial societies.

**MASON** (Multi-Agent Simulator Of Neighborhoods/Networks) is a fast discrete-event multiagent simulation library in Java, designed to be the foundation for large simulations, and also to provide functionality for lightweight simulation needs. MASON contains both a model library and an optional suite of visualization tools in 2D and 3D. It is a multiagent simulation toolkit designed to support large numbers of agents relatively efficiently on a single machine. MASON has no domain-specific features: it is neither a robotics simulator, nor a game library. Instead it belongs in the class of domain-independent simulators which may be informally described as the "dots on a screen" type.

**MASS** is designed for a large number of social or biological agents and can be used to simulate their emergent collective behavior that may be difficult only with mathematical and macroscopic approaches. A successful key for simulating mega-scale agents is to speed up the execution with parallelization. MASS has been developed to address the parallelization challenges, and a new parallel-computing library for multiagent and spatial simulation over a cluster of computing nodes has been created. MASS composes a user application of distributed arrays and agents, each representing an individual simulation place or an active entity. All computation is enclosed in each array element or agent. All communication is scheduled as periodic data exchanges among those entities, using machine-independent identifiers. Agents migrate to a remote array element for rendezvousing with each other. The agent-based approach of MASS takes advantage of these merits for parallelizing big data analysis using climate change and biological network motif searches as well as individual-based simulation such as neural network simulation and influenza epidemic simulation as practical application examples.

**Mesa** is a modular framework for building, analyzing and visualizing agent-based models. Mesa is modular, meaning that its components are kept separate but intended to work together. The modules are grouped into three categories: *Modeling*: modules used to build the models themselves: a model and agent classes, a scheduler to determine the sequence in which the agents act, and space for them to move around on; *Analysis*: tools to collect data generated from the model, or to run it multiple times with different parameter values; and *Visualization*: classes to create and launch an interactive model visualization, using a server with a JavaScript interface.

**MOOSE** (Multiphysics Object-Oriented Simulation Environment) is a finite-element, multiphysics framework which provides a high-level interface to some of the most sophisticated nonlinear solvers. It has a straightforward API that aligns well with the real-world problems scientists and engineers need to tackle. Some of its capabilities are: a multiphysics solver, dimension-independent physics, the ability to run on more than 100,000 CPU cores, a modular development simplifies code reuse, intuitive parallel multiscale solvers, dimension-agnostic, parallel geometric search, a graphical user interface and pluggable interfaces that allow the specialization of every part of the solver.

**Orleans** is a framework that provides a straightforward approach to building distributed high-scale computing applications, without the need to learn and apply complex concurrency or other scaling patterns. Project Orleans introduced the Virtual Actor abstraction, which provides a fairly simple approach to building distributed interactive applications, without the need to learn complex programming patterns for handling concurrency, fault tolerance, and resource management. Orleans applications scale-up automatically and are meant to be deployed in the Azure cloud.

**Repast** is a family of agent-based modeling and simulation platforms. "Repast Symphony" is an interactive and easy to learn Java-based modeling system that is designed for use on workstations and small computing clusters. "Repast for High Performance Computing" is a lean and expert-focused modeling system based on C++ that is designed for use on large computing clusters and supercomputers.

**SeSAm** (Shell for Simulated Agent Systems) is a generic environment for the development and simulation of multiagent models, especially when modeling societies. Moreover, it has some valuable properties, like the possibility for formulating flexible interaction between agents, multi-level interaction, adaptivity, etc. The main entities in a SeSAm model are agents, resources and the world. Their state and behavior can be implemented at specification level based on visual programming. There are also some aspects that allow scaling up for complex multiagent simulation: user functions, user features and model-specific data types. Simulation runs may be executed for different situations and aggregated into so-called experiments. Also, model instrumentation for gathering and visualizing simulation data is possible via so-called analysis. Before starting a simulation run, the model is compiled using standard optimization techniques from compiler theory, thus visual programming is combined with fast execution. Execution may also be distributed over a network using the remote simulation runs.

**SLAPP** is a simulation shell for agent systems based on a swarm-like agent protocol implemented in Python 3.

**SPADE** (Smart Python multi-agent Development Environment) is a multiagent and organizations platform based on the XMPP/Jabber technology, which offers many features that ease the construction of MAS, such as an existing communication channel, the concepts of users (agents) and servers (platforms) and an extensible communication protocol based on XML, just like FIPA-ACL. SPADE is the first to base its roots on the XMPP technology. The SPADE Agent Library is a collection of classes, functions and tools for creating agents that can work with the SPADE Agent Platform. Spade-BDI is a plugin that implements BDI agents.

**SpaDES** is an R meta-package for implementing a variety of event-based models, with a focus on spatially explicit models. These include raster-based, event-based, and agent-based models. The core simulation components are built upon a discrete event simulation framework that facilitates modularity, and easily enables the user to include additional functionality by running user-built simulation modules.

## 3.2. Commercial Software or with Unknown License Types

In this section, we include some general purpose platforms that require purchasing a commercial license, although most of them offer discounts, or free versions for academia. These platforms are very feature-rich, offer a vast array of reusable components, and can be used in a wide range of applications, with industrial strength simulations. They also offer supporting tools for model analysis and validation.

*Table 2. General purpose platforms - commercial software or with unknown license type*

| No. | Name | Programming language | Website. Projects and applications | License | Description |
|---|---|---|---|---|---|
| 1 | AnyLogic | Java and GUI based development | https://www.anylogic.com/features/ | Commercial license, free version for education | Multimethod simulation (discrete event, agent-based, system dynamics) |
| 2 | ExtendSim | ModL and GUI based. | https://extendsim.com/ | Commercial license, discounts for academia | continuous process modeling, discrete-event simulation, RBD tool, component based, hierarchical components modeling, many pre-built components |
| 3 | FlexSim | C++ and FlexScript, a C-like language, GUI based interaction | www.flexsim.com | Commercial license, free version for academia | Highly realistic 3D simulation modeling and analysis, vast array of pre-built components, graphical model analysis and statistical validation of models |
| 4 | FlexSim HC | C++ and FlexScript, a C-like language, GUI based interaction | https://healthcare.flexsim.com/healthcare-simulation-software/ | Commercial license, free version for academia | variant of the FlexSim software, designed specifically for the unique challenges faced by the healthcare facilities. |
| 5 | GoldSim | GUI | https://www.goldsim.com/Web/Home/ | Commercial license, big discount for academia | Monte Carlo simulation for dynamically modeling complex systems in engineering, science and business; supports decision-making and risk analysis |
| 6 | JACK | Java | http://aosgrp.com/products/jack/<br>*Projects:*<br>http://aosgrp.com/applications/<br>http://aosgrp.com/featured-research/ | Commercial license | BDI, plan language, graphical plans (reasoning graphs) |
| 6 | JIAC, micro JIAC, ASGARD | Java | http://www.jiac.de/<br>*Projects:*<br>http://www.jiac.de/projects/<br>*Description of the framework and tools:*<br>https://pdfs.semanticscholar.org/3267/c77443efbfbbfe027b4f94432a1a1544c9f5.pdf?_ga=2.80982996.259933356.1571138905-1270923650.1571138905 | GNU/GPL | BDI, rule engine, security; large scale distributed applications and services |
| 7 | **Simio** | GUI programming | https://www.simio.com/software/simulation-software.php | Closed source, Some free versions for academic use | supports continuous process and discrete event systems, and mixing them; real-time risk analysis; pre-built components, 3D modeling environment |
| 8 | **SIMUL8** | GUI based programming | https://www.simul8.com/ | Commercial license, some free options for academic use | Very comprehensive general purpose, discrete event, agent-based, continuous process or hybrid. |
| 9 | **Simulink** | StateFlow, Simscape | https://www.mathworks.com/products/simulink.html | Commercial license, free options for academia | large-scale models through componentization, vast array of reusable system components and libraries; massive simulations in parallel on desktop, cluster or cloud with minimal coding; hardware verification and validation |
| 10 | **Wolfram SystemModeler** | Wolfram language, C/C++, Java, Modelica language by Wolfram | www.wolfram.com/system-modeler/ | Commercial license with significant discounts for academia | General purpose modeling and simulation environment; industrial strength multi domain models of complete systems |

**AnyLogic** is a professional software tool for building industrial strength agent-based models. Simulation can be performed for marketing, social processes, and healthcare/epidemic models. ABM allows leveraging an organization's big data to populate large-scale models with agents with personalized properties such as consumer behavior, individual skills, schedules, performance data, or health-related profiles. Agent-based simulation allows representations of entities and resources as agents with individual parameters and behavior. It also supports discrete event modeling, when the process or business can be described as a sequence of events. It also supports system dynamics modeling, when individual properties of people, products, or events can be ignored.

**ExtendSim** is a family of simulation tools that support continuous process modeling and discrete-event simulation. "ExtendSim CP" is a product for modeling continuous, time-based processes. It can graphically represent the dynamics of continuous processes, exchange information with other applications, manage big data, perform analysis and report results. It has many pre-built components to model system behavior and supports unlimited levels of hierarchy for the system components. "ExtendSim DE" helps to build a comprehensive message-based discrete event architecture for intuitive modeling of any system where time advances when events occur and tracking of model entities is important. It includes pre-built discrete event blocks to represent item generation, queues, activities and delays, shutdown and shift. It has features such as: priorities, preemption, reneging, jockeying, blocking and balking. "ExtendSim Pro" offers additional features such as "Discrete Rate Module", for simulations that involve tanks, levels, valves, and modeling of the storage and rate-based movement of system components. It includes a reliability block diagramming tool.

**FlexSim** is 3D simulation software that models, simulates, predicts, and visualizes business systems in a variety of industries: manufacturing, material handling, healthcare, warehousing, mining, logistics, and more. It can model real systems in native, highly realistic and immersive 3D virtual environment, accounting for real-world variability with the help of the statistical distributions and random number generation that it supports. It contains many objects that can be used for modeling and it can also import CAD 3D drawings.

**FlexSim HC** is a variant of the FlexSim software, designed for the challenges faced by healthcare facilities. The "Patient Track" emulates the natural pull of human resources that come to the patient to provide service and guide them to the appropriate next step in their prescribed care process.

**GoldSim** is a software platform for visualizing and simulating the future behavior of physical, financial or organizational systems. A model can be built in an intuitive manner by drawing diagrams of the system. GoldSim has features to quantitatively represent the random variability and uncertainty that is present in all systems using Monte Carlo simulations. It is a hybrid simulator, allowing to superimpose the occurrence and consequences of discrete events (financial transactions, accidents, failures, etc.) onto continuously varying systems. This ability, coupled with features to support the construction of hierarchical, top-down models, facilitates the simulation of large, complex systems while keeping the models easy to understand and navigate.

**JACK** (Jack Intelligent Agent) is a framework for agent system development which uses the BDI model and provides its own Java-based plan language. These plans are compiled into Java classes for execution. It also allows the programmer to write and manage graphical plans, which represent the discrete reasoning of an agent displayed graphically as a flow chart.

**JIAC, micro JIAC, ASGARD** (Java-based Intelligent Agent Componentware) is an agent architecture and framework that eases the development and the operation of large-scale, distributed applications and services. The framework supports the design, implementation, and deployment of software agent systems. It includes: BDI, rule engine, security. JIAC is the framework, ASGARD is a tool meant as a monitor for controlling distributed multiagent infrastructure at runtime. Micro JIAC is the lightweight version for constrained devices.

**Simio** Simulation Software provides an object-based 3D modeling environment which allows the construction of a 3D model in a single step, from a top-down 2D view. All Simio model-building products integrate with Google Warehouse to allow access to a library of freely available 3D symbols that can add realism to the models. Simio simulation software was used for manufacturing, healthcare, aerospace and defense, mining, industrial engineering. It can model continuous and discrete systems.

**SIMUL8** supports simulations for continuous processes, discrete events, or agent-based models for various domains. It also supports optimization through the "OptQuest" plugin, and "Simul8 Online" is an online version, with the same feature set.

**Simulink** is a very complex software suite from Mathworks, used to perform multi domain modeling and simulation, because models can be reused across environments to simulate how all parts of the system work together. Hybrid systems can be built and simulated. With "Stateflow" one can model combinatorial and sequential logic with state machines and flowcharts, and "SimEvents" can be used to represent agents and event-driven processes. "Stateflow" provides a graphical language that includes state transition diagrams, flow charts, state transition tables, and truth tables. Physical systems can be modeled using "Simscape", which supports C-code generation to deploy models to other simulation environments, including hardware-in-the-loop (HIL) systems. "SimEvents" enables the study of the effects of task timing and resource usage on the performance of distributed control systems, software and hardware architectures, and communication networks.

**Wolfram SystemModeler** is a modeling and simulation environment for cyber-physical systems. Using drag and drop from the built-in modeling libraries, one can build industrial strength, multidomain models of complete systems. The Wolfram Language can be used for analyzing, understanding and quickly iterating system designs.

## 4. Special Purpose Platforms

### 4.1. Cognitive, Social and Affective Agent Platforms

The platforms included in this section can be used to model human behavior and social phenomena. Some of them are only in the phase of framework design, without an actual implementation.

*Table 3. Cognitive, Social and Affective Agent Platforms*

| No. | Name | Programming language | Website. Projects and applications | License | Description |
|---|---|---|---|---|---|
| 1 | **ABC-EBDI** | No code | https://www.sciencedirect.com/science/article/pii/S1389041718309136 | No product, just framework design | Affective, BDI, human behavior, cognitive |
| 2 | **ACT-R** | ACT-R | http://act-r.psy.cmu.edu/ <br> *Projects:* <br> http://acs.ist.psu.edu/papers/ritterTOip.pdf | Open source (Lisp) | Human behavior, cognitive agents |
| 3 | **Cormas** | Smalltalk | http://cormas.cirad.fr/indexeng.htm | Free, closed source | Relationships between societies and their environments, interactive simulation |
| 4 | **EBDI** | No code | https://www.researchgate.net/publication/221454893_EBDI_an_architecture_for_emotional_agents | No product, just framework/architecture design | BDI, emotional agents, merge of emotion theories with agent reasoning process. |
| 5 | **DALI** | Prolog | https://github.com/AAAI-DISIM-UnivAQ/DALI <br><br> https://www.aaai.org/Papers/Symposia/Spring/2008/SS-08-02/SS08-02-003.pdf | Apache License 2.0 | An architecture and an agent-oriented logic programming language |
| 6 | **GOAL** | GOAL It currently uses Prolog as a knowledge representation language | https://goalapl.atlassian.net/wiki/spaces/GOAL/overview?mode=global <br> *Projects:* <br> https://goalapl.atlassian.net/wiki/spaces/GOAL/pages/33046/Projects | Open source (Java) | GOAL allows and facilitates the manipulation of an agent's beliefs and goals and to structure its decision-making |
| 7 | **GROWLab** | Java | https://icr.ethz.ch/research/growlab/ | Open source (Java) | Social phenomena, hierarchical relationships between model actors. |
| 8 | **Jason** | AgentSpeak | http://jason.sourceforge.net/ <br> *Projects:* <br> http://jason.sourceforge.net/wp/projects/ | | fully fledged interpreter for AgentSpeak, BDI |
| 9 | **Soar** | Soar | http://soar.eecs.umich.edu/ <br> *Projects:* <br> http://soar.eecs.umich.edu/groups <br> https://en.wikipedia.org/wiki/Soar_(cognitive_architecture) | Open source (C++) | Cognitive architecture |
| 10 | SOSIEL | C# | https://www.sosiel.org<br>*Github:*<br>https://github.com/SOSIEL/Platform-SOSIEL | Open source (C#) | Agent system simulating social learning, collective action, cross-generational population dynamics, multi-layered social network structures; |
| 11 | The Matrix | No code provided | http://www.ifaamas.org/Proceedings/aamas2019/pdfs/p1635.pdf | No code/product offered, just framework design | Data-intensive simulation at-scale. |

**ABC-EBDI** is an "affective" framework for BDI agents. The aim of this framework is to improve the modeling of intelligent agents that reproduce realistic human behavior. BDI frameworks have been successfully used in this context and have evolved in recent years into EBDI frameworks, in which affective aspects are considered. The distinguishing feature of this framework is the classification of an agent's cognitive-affective process as either rational or irrational. This process leads to functional emotions and adaptive conduct in the first case and dysfunctional emotions and maladaptive behaviors in the second. The framework models affect by considering emotions, mood and personality. It also models human conduct regarding not only actions, but also the way those actions are expressed.

**ACT-R** is a cognitive architecture: a theory about how human cognition works. On the exterior, ACT-R looks like a programming language; however, its constructs reflect assumptions about human cognition. These assumptions are based on numerous facts derived from psychological experiments. These assumptions can be tested by comparing the results of the model (i.e. the traditional measures of cognitive psychology: time to perform the task, accuracy in the task, or, more recently, neurological data such as those obtained from FMRI) with the results of people doing the same tasks. One important feature of ACT-R that distinguishes it from other theories in the field is that it allows researchers to collect quantitative measures that can be directly compared with the quantitative measures obtained from human participants.

**Cormas** is a generic ABM platform dedicated to common-pool resource management which can be used to understand the relationships between societies and their environment. It is intended to facilitate the design of ABM as well as the monitoring and analysis of simulation scenarios. Cormas can enable stakeholders to interact with the execution of a simulation by modifying the behavior of the agents and the way they use the resources. It is a simulation platform based on the VisualWorks programming environment which allows the development of applications in the Smalltalk object-oriented language. Its predefined entities are Smalltalk generic classes from which, by specialization and refining, users can create specific entities for their own model. It facilitates the construction of agent-based models and the design, monitoring and analyzing of agent-based simulation scenarios. Cormas was primarily oriented towards the representation of interactions between stakeholders about the use of natural renewable resources.

**DALI** is a meta-interpreter built on top of Prolog. It has a logic programming language for modeling agents and agent systems in computational logic. The basic objective of the specification of this language is the identification and the formalization of the basic patterns for reactivity, proactivity, internal "thinking" and "memory".

**EBDI** is an architecture for emotional agents. Most of the research on agent systems has focused on the development of rational utility-maximizing agents. However, research shows that emotions have a strong effect on people's physical states, motivations, beliefs, and desires. By introducing primary and secondary emotion into BDI architecture, this framework shows a generic architecture for an emotional agent, EBDI, which can merge various emotion theories with an agent's reasoning process. It implements practical reasoning techniques separately from the specific emotion mechanism. The separation allows the user to plug in emotional models as needed or upgrade the agent's reasoning engine independently.

**GOAL** is another language for programming cognitive agents. GOAL agents derive their choice of action from their beliefs and goals. The language provides the key building blocks for designing and implementing cognitive agents. It allows and facilitates the manipulation of an agent's beliefs and goals and to structure its decision-making. GOAL is a rule-based programming language. A GOAL agent program consists of six different sections, including the knowledge, beliefs, goals, action rules, action specifications, and percept rules, respectively. The knowledge, beliefs and goals are represented in a knowledge representation language such as Prolog, Answer set programming, SQL (or Datalog), or the Planning Domain Definition Language, for example. The distinguishing feature of GOAL is the concept of a declarative goal. The goals of an agent describe what it wants to achieve, not how to achieve that.

**GROWLab** was designed to facilitate the modeling, simulation, analysis, and validation of complex social processes, with a special focus on geographic conflict research. Four core support: the seeding of the model with empirical facts (including geo-referenced data) to calibrate the environments and mechanism to the appropriate level of realism; the effective modeling of complex network and hierarchical relationships between model actors and the efficient scheduling of their interactions; the execution of large number of simulation runs on a grid made of many independent computers to test the sensitivity of the models; the statistical and visual analysis of the state of the system, as well as the unfolding of the processes over time.

**Jason** is a platform for the development of agent systems. AgentSpeak has been one of the most influential abstract languages based on the BDI architecture. The agents programmed with AgentSpeak are sometimes referred to as reactive planning systems. Jason is the first fully-fledged interpreter for an improved version of AgentSpeak, including also speech-act based inter-agent communication.

**Soar** is a general cognitive architecture for developing systems that exhibit intelligent behavior. The Soar Markup Language allows agents to communicate with external environments and it can be used to interface with agents written in other languages. The goal of the Soar project is to develop the fixed computational building blocks necessary for general intelligent agents – agents that can perform a wide range of tasks and encode, use, and learn all types of knowledge to realize the full range of cognitive capabilities found in humans, such as decision making, problem solving, planning, and natural language understanding. It is both a theory of what cognition is and a computational implementation of that theory. Since its beginnings in 1983, it has been used by AI researchers to create intelligent agents and cognitive models of different aspects of human behavior. The most current and comprehensive description of Soar is (Laird, 2012).

**SOSIEL** is an agent system platform developed for building models that are capable of capturing the spatio-temporal complexity of social contexts in which the heterogeneity of knowledge, the need for learning, and the potential for collective action plays a significant role. It can simulate the cross-generational progression of one or a large number of boundedly-rational agents, each of which is represented by a cognitive architecture that consists of theoretically-grounded cognitive processes and agent-specific and empirically-grounded knowledge. The agents can interact among themselves and with coupled natural or technical systems, learn from their and each other's experience, create new practices, make decisions take potentially collective actions.

Highlights of the platform include: an agent system simulating social learning, collective action, cross-generational population dynamics, and the self-organization of multi-layered social network structures; agent cognition is represented with a general cognitive architecture that consists of a memory component, a learning component, and a decision-making component and that can be set to one of four cognitive levels; agents can have place-based and hypothetical knowledge that is organized, updated, modified and utilized by the cognitive architecture.

**The Matrix** addresses data-intensive simulation at-scale. It can model 3 million users (each as an individual agent), 13 million repositories and 239 million user-repository interactions on GitHub. Simulations predict user interactions with GitHub repositories. They demonstrate a three-order of magnitude increase in the number of cognitive agents simultaneously interacting.

### 4.2. Platforms for Artificial Intelligence Research

The platforms included in this section are intensively used in the domain of artificial intelligence research, most of them leveraged using the Python programming language. These platforms provide various "environments" with which an intelligent agent can interact, and in various research domains such as games, autonomous driving or robotics. Some of them implement the environments using high-fidelity physics engines.

*Table 4. Platforms for artificial intelligence research*

| No. | Name | Programming language | Website. Projects and applications | License | Description |
|---|---|---|---|---|---|
| 1 | **Coach** | Python | https://nervanasystems.github.io/coach/ | Open source (Python) | AI research in autonomous driving, robotics, games, etc.; models the interaction between an intelligent agent and various environments |
| 2 | **DeepMind Garage** | Python, Tensorflow, Pytorch | https://github.com/rlworkgroup/garage | Open source (Python) | Intelligent agents development, using simulations in a variety of environments (games, control tasks, etc.) |
| 3 | **DeepMind Lab** | Python | https://github.com/deepmind/lab | Open source (C, Lua, Python) | 3D navigation and puzzle-solving environments for intelligent agent experimentation (especially deep reinforcement learning) |
| 4 | **DeepMind Spriteworld** | Python | https://deepmind.com/research/open-source/spriteworld *Github:* https://github.com/deepmind/spriteworld/tree/master/spriteworld | Open source | 2D arena with shapes that can be moved freely; small scale experiments for limited computational resources; experimentation with reinforcement learning agent implementations |
| 6 | **GAZEBO** | JavaScript | http://gazebosim.org/ | Open source | 3D robot populations simulation, indoor and outdoor; very high-fidelity physics, high-quality graphics, pre-built robot models and environments |
| 7 | **MADP** | C++ | http://www.fransoliehoek.net/fb/index.php?fuseaction=software.madp<br><br>*Github:*<br>https://github.cm/MADPToolbox/MADP | Open source | Research in decision planning and learning in agent systems |
| 8 | **MAgent** | Python | https://github.com/geek-ai/MAgent<br>https://arxiv.org/pdf/1712.00600.pdf | Open source | Supports huge number of agents; multiagent reinforcement learning |
| 9 | **MuJoCo** | C/C++ | http://www.mujoco.org/index.html | Commercial license, free student edition, special licenses for academia | first full-featured simulator designed for the purpose of model-based optimization, and in particular optimization through contacts; robotics, biomechanics, graphics, animation; 3D visualization |
| 10 | **Neural-MMO** | Python | https://github.com/jsuarez5341/neural-mmo | Open source (Python) | Massively multiagent game environment for training and evaluating intelligent agents |
| 11 | **OpenAI Gym** | Python | https://gym.openai.com/<br>*agents' baselines:*<br>https://github.com/openai/baselines | Open source | Reinforcement learning environments simulations; control tasks, Atari games emulators that allow custom agents to play in and try to solve them |
| 12 | **OpenSpiel** | C++/Python | https://github.com/deepmind/open_spiel | Open source (C++, Python, Swift) | Environments for n-player zero-sum, cooperative and general-sum, one-shot and sequential, strictly turn-taking and simultaneous-move, perfect and imperfect information games, as well as traditional multiagent |
| 13 | **PySC2 – Starcraft II Learning Environment** | Python | https://github.com/deepmind/pysc2 | Open source | AI research in the real-time strategy game StarCraft II |
| 14 | **Unity ML Agents** | Python | https://github.com/Unity-Technologies/ml-agents | Open source | AI research; 2D, 3D, VR/AR games; high fidelity physics and graphics; offers implementation of some of the best RL agents; multiagent environments |

**Coach** is a Python framework which models the interaction between an agent and an environment in a modular way. It allows one to model an agent by combining various building blocks and training the agent in multiple environments. The available environments allow testing the agent in different fields such as robotics, autonomous driving, games and more. It exposes a set of APIs for experimenting with new reinforcement learning (RL) algorithms and allows the integration of new environments to solve. Coach collects statistics from the training process and supports visualization techniques for debugging the agent being trained. It also contains the implementation of many state-of-the-art algorithms.

**DeepMind Garage** is a framework for developing and evaluating reinforcement learning algorithms in a variety of environments in which simulations for training and testing can be done. The toolkit provides a wide range of modular tools for implementing RL algorithms, as well as the implementation of some of the best RL algorithms. Custom agents and environments can be created.

**DeepMind Lab** provides a suite of challenging, customizable, 3D navigation and puzzle-solving tasks for learning agents. Its primary purpose is to act as a testbed for research in artificial intelligence, especially deep reinforcement learning.

**DeepMind Spriteworld** is a Python-based RL environment that consists of a 2-dimensional arena with simple shapes that can be moved freely. Spriteworld is suited for small-scale experiments with limited computational resources. Spriteworld sprites can have many shapes and can vary continuously in position, size, color, angle, and velocity. The environment has occlusion but no physics, so by default sprites pass beneath each other but do not collide or interact in any way. Interactions may be introduced through the action space, which can update all sprites each timestep. There are a variety of action spaces, some of which are continuous (like a touch-screen) and others of which are discrete (like an embodied agent that takes discrete steps).

**GAZEBO** is a 3D dynamic simulator with the ability to accurately and efficiently simulate populations of robots in complex indoor and outdoor environments. While similar to game engines, Gazebo offers physics simulation at a much higher degree of fidelity, a suite of sensors, and interfaces for both users and programs. Typical uses of Gazebo include testing robotics algorithms, designing robots, performing regression testing with realistic scenarios. A few key features of Gazebo include: multiple physics engines, a library of robot models and environments, a variety of sensors, and programmatic and graphical interfaces.

**MADP** (Multiagent Decision Process) is a software toolbox for scientific research in decision-theoretic planning and learning in agent systems. MADP refers to a collection of mathematical models for multiagent decision making: multiagent Markov decision processes (MMDPs), decentralized MDPs, decentralized partially observable MDPs, and partially observable stochastic games, etc. It provides classes modeling the basic data types used in MADPs (e.g., action, observations, etc.) as well as derived types for planning (observation histories, policies, etc.). It also provides base classes for planning algorithms and includes several example applications using the provided functionality. Several utility applications are provided, for instance one which empirically determines a joint policy's control quality by simulation.

**MAgent** is a research platform for many-agent reinforcement learning. It aims at supporting RL scenarios that scale up from hundreds to millions of agents.

**MuJoCo** is a physics engine aiming to facilitate research and development in robotics, biomechanics, graphics and animation, and other areas where fast and accurate simulation is needed. It is designed for the purpose of model-based optimization, especially through contacts. MuJoCo makes it possible to scale up computationally intensive techniques such optimal control, physically-consistent state estimation, system identification and automated mechanism design, and apply them to complex dynamical systems in contact-rich behaviors. It also has more traditional applications such as testing and validation of control schemes before deployment on physical robots, interactive scientific visualization, virtual environments, animation and gaming. Its key features are: simulation in generalized coordinates, avoiding joint violations; inverse dynamics that are well-defined even in the presence of contacts; unified continuous-time formulation of constraints via convex optimization; constraints include soft contacts, limits, dry friction, equality constraints; • simulation of particle systems, cloth, rope and soft objects; actuators including motors, cylinders, muscles, tendons, slider-cranks; Newton, Conjugate Gradient, or Projected Gauss-Seidel solvers; pyramidal or elliptic friction cones, dense or sparse Jacobians; Euler or Runge-Kutta numerical integrators; multi-threaded sampling and finite-difference approximations; cross-platform GUI with interactive 3D visualization in OpenGL; run-time module written in C and tuned for performance.

**Neural-MMO** platform supports a large, variable number of agents within a persistent and open-ended task. The inclusion of many agents and species leads to better exploration, divergent niche formation, and greater overall competence. The platform satisfies the following criteria: Persistence: agents learn concurrently in the presence of other learning agents with no environment resets. Strategies must consider long time horizons and adapt to potentially rapid changes in the behaviors of other agents; Scale: the environment supports a large and variable number of entities, e.g. up to 100M lifetimes of 128 concurrent agents in each of 100 concurrent servers; Efficiency: effective policies can be trained on a single desktop CPU; Expansion: core features include procedural generation of tile-based terrain, a food and water foraging system, and a strategic combat system.

**OpenAI Gym** library contains a collection of environments that can be used to test the intelligent agents' ability to solve them. Environments are grouped as follows: classic control, which are small-scale control tasks like a small robot arm, pendulum, etc.; simple 2D and 3D robots, which are used by intelligent agents that are trying to control a robot in simulation. These environments are based on MuJoCo engine; Atari games suite (https://github.com/mgbellemare/Arcade-Learning-Environment) which allows experimentation of reinforcement learning algorithms, by running simulations of reinforcement learning agents; algorithmic, where the challenge is to learn the algorithms from examples.

**OpenSpiel** is a collection of environments and algorithms for research in general reinforcement learning and search or planning in games. OpenSpiel supports *n*-player single- and multiagent zero-sum, cooperative and general-sum, one-shot and sequential, strictly turn-taking and simultaneous-move, perfect and imperfect information games, as well as traditional multiagent environments such as partially- and fully-observable grid worlds and social dilemmas. It also includes tools to analyze learning dynamics and common evaluation metrics. The core API and games are implemented in C++ and exposed to Python. Algorithms and tools are written both in C++ and Python.

**PySC2 Starcraft II Learning Environment** is DeepMind's Python component of the StarCraft II Learning Environment (SC2LE). It exposes Blizzard Entertainment's StarCraft II Machine Learning API as a Python RL environment. PySC2 provides an interface for RL agents to interact with StarCraft 2, getting observations and sending actions. This environment wrapper helps by offering an interface for RL agents to play the game. The game is broken down into "feature layers", where the elements of the game such as unit type, health and map visibility are isolated from each other, while preserving the core visual and spatial elements of the game.

**Unity ML-Agents** (The Unity Machine Learning Agents Toolkit) is an open-source Unity plugin that enables games and simulations to serve as environments for training intelligent agents. Agents can be trained using reinforcement learning, imitation learning, neuroevolution, or other machine learning methods through a Python API. This toolkit provides implementations (based on TensorFlow) of state-of-the-art algorithms to enable users to train intelligent agents for 2D, 3D and VR/AR games.

### 4.3. Platforms for Modeling and Simulating Environments and Ecosystems

This section presents some platforms that can be used for modeling and simulating natural phenomena, with environmental, climate, hydrological, wildlife or agricultural applications.

*Table 5. Platforms for modeling and simulating environments and ecosystems*

| No. | Name | Programming language | Website. Projects and applications | License | Description |
|---|---|---|---|---|---|
| 1 | Agent Analyst | *NQPy - Not Quite Python*, a subset of Python. | http://resources.arcgis.com/en/help/agent-analyst/ | Open source | Agent based models with spatial components based on ArcGIS |
| 2 | Altreva Adaptive Modeler | UI based | https://www.altreva.com/product.htm | Commercial license, free evaluation | Real-time price forecasting in real world markets |
| 3 | BSim | Java | https://cellsimulationlabs.github.io/tools/bsim/index.html<br><br>*Published paper:*<br>https://pubs.acs.org/doi/pdf/10.1021/acssynbio.7b00121<br><br>*Github:*<br>https://github.com/CellSimulationLabs/bsim | Open source (Java) | Bacterial population study, Realistic 3D environments, accurate physics, built-in models |
| 4 | CRAFTY | XML | https://www.wiki.ed.ac.uk/display/CRAFTY/Home | Open source (Java) | Lands use and goods and services they produce, large scale |
| 5 | EMOD | JSON based modeling, Python runner scripts | http://www.idmod.org/software#emod<br>https://github.com/InstituteforDiseaseModeling/EMOD | Open source (C++) | Quantitative and analytical means to model infectious disease |
| 6 | Envision | Xml based specification | http://envision.bioe.orst.edu/ | Open Source (C++) | spatially-explicit modeling platform specifically designed for scenario-based exploration of coupled human and natural systems (CHANS), GIS based |
| 7 | Framsticks | FramScript (similar to JavaScript, Java), C++ | http://www.framsticks.com/<br>http://www.framsticks.com/wiki/Projects.html<br>http://www.framsticks.com/dev/main.html | Open source (C++) | three-dimensional life simulation |
| 8 | HexSim | Modeling done through GUI, but engine written in C++ and the GUI in C# | http://www.hexsim.net/ | Free, closed source, C++/C#. soon to be made open source | Spatially explicit, individual based, multi-species, life history simulator. |
| 9 | **iLand** | Javascript | http://iland.boku.ac.at/startpage | GPL, open source (C++) | General model of forest ecosystem dynamics; resilience of ecosystems, climate change adaptation, forest carbon storage and exchange, functional roles of diversity; spatially explicit, hierarchical multi-scale |
| 10 | **LANDIS-II** | C# | http://www.landis-ii.org/ <br> *Projects*: <br> http://www.landis-ii.org/projects <br> https://github.com/LANDIS-II-Foundation | Open source (C#) | Forest landscape model, multi-century timescales, spatial scales spanning hundreds of millions of hectares. |
| 11 | **MIRAGE** | No code provided | http://www.geocomputation.org/2015/papers/GC15_56.pdf | No code/product provided, just design description | High performance computing agent-based framework that can capture dynamics and behaviors of billions of humans. |
| 12 | **MPMAS** | Excel files | https://mp-mas.uni-hohenheim.de/ | Free, closed source (C++) | Large-scale simulations of land use change in agriculture and forestry; farm-level linear programming |
| 13 | **NetLogo** | Logo dialect | https://ccl.northwestern.edu/netlogo/ <br> *Model library*: <br> http://ccl.northwestern.edu/netlogo/models/ <br> *MAS models*: <br> http://jmvidal.cse.sc.edu/netlogomas/ | | programmable modeling environment for simulating natural and social phenomena |
| 14 | **OMS** - Object modeling system | Java | https://alm.engr.colostate.edu/cb/wiki/16961 | Open source (Java) | Distributed hydrological and environmental models |
| 15 | **Osmose** | R (pre/post processing tools, call to the JAVA core, etc.) and Java | http://www.osmose-model.org/object-oriented-simulator-marine-ecosystems <br> *Code*: <br> https://github.com/osmose-model/osmose | Open source (Java and R) | Object-oriented simulator of marine ecosystems |
| 16 | **TerraME** | Lua | http://www.terrame.org/doku.php?id=start | Open source (Lua, C++) | Spatial dynamical modeling; cellular automata, agent-based models, network models running in 2D cell spaces; access to TerraLib geographical database; Anisotropic spaces and hybrid automata models. |
| 17 | **UrbanSim** | GUI based programming | https://urbansim.com/ | Commercial license | Simulation platform for supporting planning and analysis of urban development, incorporating the interactions between land use, transportation, the economy, and the environment; 3D visualizations |

**Agent Analyst** is a platform for agent-based modeling which integrates Esri ArcGIS, a geographic information system, and Repast agent system. It can be used to simulate, for example, the behavior of wildlife directly on a geographic map.

**Altreva Adaptive Modeler** creates agent-based models for price forecasting of real-world markets such as stocks, cryptocurrencies, ETFs, commodities or forex currency pairs. One-step-ahead forecasts and trading signals are generated after every received price bar or tick. Models are created mostly automatically. Some parameters can be adjusted by the user to take into account the specific characteristics of the security such as the data fields to use (i.e. open, high, low close, bid, ask, volume), minimum price increment, transaction costs and the user's trading preferences. It is also possible to experiment with agent-based model settings such as the number of market participants (agents), their initial wealth and asset distribution, and evolution settings. Model evolution is visualized in real-time and can be paused and resumed by the user at any time. It is also possible to advance the model step-by-step. "Adaptive Modeler" contains a "Trading Simulator" to simulate trading based on the trading signals.

**BSim** is an agent-based modeling tool designed to allow for the study of bacterial populations. By enabling the description of bacterial behaviors, it attempts to provide an environment in which to investigate how local interactions between individual bacteria leads to the emergence of population level features, such as cooperation and synchronization.

**CRAFTY** is a large-scale ABM designed to simulate scenarios related to land uses and the goods and services they produce. It can be used without the need of programming.

**EMOD** (Epidemiological MODeling software) simulates the spread of disease to help determine the combination of health policies and intervention strategies that can lead to disease eradication. EMOD is a stochastic, mechanistic, agent-based model that simulates the actions and interactions of individuals within geographic areas to understand the disease dynamics in a population over time. EMOD supports modeling a variety of different diseases including malaria, HIV, and tuberculosis. IDM has created a suite of over 600 scenarios. A vital part of this suite of tests is the "scientific feature testing," which involves validating the model output with the mathematical formulas and distributions specified for each disease.

**Envision** is a robust, spatially-explicit modeling platform specifically designed for scenario-based exploration of coupled human and natural systems. It can integrate traditional simulation models with a multiagent modeling system, incorporating actors and policies capturing the decision rules available to those actors.

**Framsticks** is a three-dimensional life simulation project. Both mechanical structures ("bodies") and control systems ("brains") of creatures are modeled. It is possible to design various kinds of experiments, including simple optimization (by evolutionary algorithms), coevolution, open-ended and spontaneous evolution, distinct gene pools and populations, diverse genotype-phenotype mappings, and modeling of species and ecosystems. The system can be interesting for experimenters who would like to evolve their own artificial creatures and see them in a three-dimensional, virtual world.

**HexSim** is a spatially explicit, individual-based computer model designed for simulating terrestrial wildlife population dynamics and interactions. It is a framework within which plant and animal population models are constructed. Users define the model structure, complexity, and data needs. Every function can be accessed through a graphical user interface (GUI). HexSim uses spatial data to capture landscape structure, habitat quality, stressor distribution, and other types of information. It can be employed for exploring the cumulative impact to wildlife populations and plants resulting from multiple interacting stressors. HexSim simulations are built around a user-defined life cycle, which is the principal mechanism driving all other model processing and data needs. The life cycle consists of a sequence of life-history events that are selected from a list, which includes survival, reproduction, movement, resource acquisition, species interactions, etc. Through the use of events, the user can impose yearly, seasonal, daily, or other temporal cycles on the simulated population.

**iLand** is a general model of forest ecosystem dynamics. It can be employed to elucidate a wide variety of ecology and management-related questions, simulating individual tree competition, growth, mortality, and regeneration. It addresses interactions between climate (change), disturbance regimes, vegetation dynamics, and forest management.

**LANDIS-II** forest landscape model simulates future forests (both trees and shrubs) at decadal to multi-century time scales and spatial scales spanning hundreds to millions of hectares. The model simulates change as a function of growth and succession and, optionally, the influence of disturbances (e.g., fire, wind, insects), forest management or land use change. LANDIS-II also provides dozens of libraries ("extensions").

**MIRAGE** is a framework for data-driven collaborative high-resolution simulation. Information about how human populations shift in response to various stimuli is limited because no single model can address these stimuli simultaneously, and integration of the best existing models has been challenging because of the vast disparity among constituent model purpose, architecture, scale and execution. To demonstrate a potential model coupling for approaching this problem, three major model components are integrated into a fully coupled system that executes a worldwide infection-infected routine where a human population requires a food source for sustenance and an infected population can spread the infection when they are in contact with the remaining healthy population.

**MPMAS** (Mathematical Programming-based Multi-Agent Systems) is a software package for simulating land use change in agriculture and forestry. It combines economic models of farm household decision-making with a range of biophysical models simulating the crop yield response to changes in the crop water supply and changes in soil nutrients. MPMAS is part of a family of models called agent systems models of land-use/cover change, which couple a cellular component representing a physical landscape with an agent-based component representing land-use decision-making. The main difference between MPMAS and alternative packages is the use of whole farm mathematical programming to simulate land-use decision-making. MPMAS was applied in a dozen countries around the world. It is flexible in terms of the spatial extent that it can cover and has been used in small-scale as well as large-scale applications.

**NetLogo** is a programmable modeling environment for simulating natural and social phenomena. It is particularly well suited for modeling complex systems developing over time. Modelers can give instructions to hundreds or thousands of agents all operating independently. This makes it possible to explore the connection between the micro-level behavior of individuals and the macro-level patterns that emerge from their interaction

**OMS** (Object Modeling System) is a Java modeling framework, which allows model construction and model application based on components. This is a collaborative project active among the U.S. Department of Agriculture and partner agencies and organizations involved with agro-environmental modeling.

**Osmose** is a multispecies and individual-based model which focuses on fish species. It assumes opportunistic predation-based on spatial co-occurrence and size adequacy between a predator and its prey. It represents fish individuals grouped into schools, which are characterized by their size, weight, age, taxonomy and geographical location (2D model), and which undergo major processes of fish life cycle (growth, explicit predation, natural and starvation mortalities, reproduction and migration) and a fishing mortality distinct for each species.

**TerraME** is a programming environment for spatial dynamical modeling. It supports cellular automata, agent-based models, and network models running in 2D cell spaces. TerraME provides an interface to TerraLib geographical database, allowing models direct access to geospatial data. Two important innovations in TerraME are its use of anisotropic spaces and of hybrid automata models. Anisotropic spaces arise when modeling natural and human-related phenomena. For example, land settlers in a new area do not occupy all places at the same time; they follow roads and rivers, leading to an anisotropic pattern. Anisotropic spaces are implemented in TerraME using generalized proximity matrices. A hybrid automaton is an abstract model for a system whose behavior has discrete and continuous parts. It extends the idea of finite automata to allow continuous change to take place between transitions. Adopting hybrid automata in spatial dynamical models allows complex models which include critical transitions.

**UrbanSim** is a simulation system for supporting planning and analysis of urban development, incorporating the interactions between land use, transportation, the economy, and the environment. It is designed for use by metropolitan planning organizations, cities, counties, non-governmental organizations, real estate professionals, planners, researchers and students interested in exploring the effects of infrastructure and development constraints as well as other policies on community outcomes such as motorized and non-motorized accessibility, housing affordability, greenhouse gas emissions, and the protection of open space and environmentally sensitive habitats. UrbanSim is a computational representation of metropolitan real estate markets interacting with transport markets, modeling the choices made by households, businesses, and real estate developers, and how these are influenced by governmental policies and investments.

## 4.4. Platforms for Transport-Related Simulations

This section presents platforms used for traffic simulations in autonomous driving scenarios, agent-based transportation simulations, as well as for autonomous vehicle simulations.

*Table 6. Platforms for Transport-Related Simulations*

| No. | Name | Programming language | Website. Projects and applications | License | Description |
|---|---|---|---|---|---|
| 1 | Carla | Python | http://carla.org/ | Open source (C++) | Autonomous driving systems, realistic 3D simulation, physics engine, traffic scenarios simulation |
| 2 | Distributed Real-Time Traffic Simulation for Autonomous Vehicle Testing in Urban Environments | No code provided | https://ieeexplore.ieee.org/document/8569544 | No product, just framework design | Traffic simulation for autonomous driving in urban environments |
| 3 | MATSim | Java | http://www.matsim.org/ Code: https://github.com/matsim-org/matsim | Open source (Java) | Large-scale agent-based transport simulation |

| 4 | Microsoft AirSim | C++, Python, C#, Java | https://github.com/microsoft/AirSim | Open Source (C++) | Simulator for autonomous vehicles built on Unreal Engine and Unity engine |
|---|---|---|---|---|---|
| 5 | Spice | N/A | https://link.springer.com/article/10.1007%2Fs10458-018-9383-2 | No code/product offered, just framework design | cognitive agent framework for computational crowd simulations in complex environments |
| 6 | Torcs - The Open Racing Car Simulator | C/C++ | http://torcs.sourceforge.net/index.php | Open source (C/C++) | Car racing simulation; realistic 3D graphics; real-time; AI racing game – research platform |

**Carla** was created to support development, training, and validation of autonomous driving systems. In addition to open-source code and protocols, CARLA provides open digital assets such as: urban layouts, buildings and vehicles. The simulation platform supports flexible specification of sensor suites, environmental conditions, full control of all static and dynamic actors, and map generation. CARLA consists mainly of two modules, the CARLA Simulator and the CARLA Python API module. The simulator controls the logic, physics, and rendering of all the actors and sensors in the scene; it requires a machine with a dedicated GPU to run. The CARLA Python API is a module that can be imported in Python scripts, it provides an interface for controlling the simulator and retrieving data. With this Python API one can control any vehicle in the simulation, attach sensors to it, and read back the data these sensors generate.

**Distributed Real-Time Traffic Simulation for Autonomous Vehicle Testing in Urban Environments** presents a distributed real-time simulation setup for automated driving function testing in urban environments. In the automotive domain, many simulation frameworks are utilized which are tailored towards a specific application. However, virtual testing of automated driving functions requires a holistic simulation of realistic urban traffic environments. This paper presents a distributed simulation framework, with integrated ego-vehicle and a linked pedestrian simulator. It also presents a pedestrian behavior model, which can interact with all agents of the different simulation instances.

**MATSim** provides a toolbox to run and implement large-scale agent-based transport simulations. The toolbox consists of several modules which can be combined or used stand-alone. Modules can be replaced by custom implementations to test single aspects of the modeler's work. It offers a toolbox for demand-modeling, agent-based mobility-simulation (traffic flow simulation), re-planning, a controller to iteratively run simulations as well as methods to analyze the output generated by the modules.

**Microsoft AirSim** is a simulator for drones, cars, etc., built on the Unreal engine (https://www.unrealengine.com, also with an experimental Unity release: https://unity.com). It is open-source, cross-platform and supports HIL with flight controllers for physically and visually realistic simulations. The goal of AirSim is to become a platform for AI research to experiment with deep learning, computer vision and reinforcement learning algorithms for autonomous vehicles. For this purpose, it also exposes APIs to retrieve data and control vehicles in a platform-independent way. It has support for multiple types of sensors, such as GPS, barometer, distance sensor, Lidar, and has multiple vehicle capability.

**Spice.** Pedestrian behavior is an omnipresent topic, but the underlying cognitive processes and the various influences on movement behavior are still not fully understood. Nonetheless, computational simulations that predict crowd behavior are essential for safety, economics, and transport. Spice provides an approach to structure pedestrian agent models by integrating concepts of pedestrian dynamics and cognition. The model solves spatial sequential choice problems in enough detail, including movement and cognition aspects.

**Torcs - The Open Racing Car Simulator** is a portable multi-platform car racing simulation. It is used as an ordinary car racing game, as AI racing game and as a research platform. It is built using a sophisticated physical model, supports many racing tracks, opponents, cars and many input devices: steering wheels, joysticks, game pads, etc.

## 5. Platforms with Unclear Status or No Longer Under Development

In this section, we present some agent based modeling and simulation platforms whose development seems to have stopped. In the following tables, the year when the latest reference (LR) was found about a specific platform (approximately), is also included. Since there are many such cases, we divide the mentioned platforms into three groups, by the decade of their latest reference: the 1990s, the 2000s and the 2010s.

Table 7 shows platforms that were built in the 1990s. Some of them are programmed using agent-oriented programming.

*Table 7. Platforms with unclear status or no longer under development (the 1990s)*

| No. | Name | LR | Website |
|-----|------|-----|---------|
| 1 | AGENT-0 | 1991 | http://infolab.stanford.edu/pub/cstr/reports/cs/tr/91/1389/CS-TR-91-1389.pdf |
| 2 | Agent-K | 1994 | http://citeseerx.ist.psu.edu/viewdoc/summary?doi=10.1.1.30.664 |
| 3 | April | 1995 | http://www.doc.ic.ac.uk/~klc/april1.html |
| 4 | OASIS | 1995 | http://www.stern.nyu.edu/om/faculty/pinedo/book2/downloads/CMU-Salman/Reference%20Articles/oasis%20aircraft%20sequencing.html |
| 5 | Agent Building Shell | 1996 | http://dl.acm.org/citation.cfm?id=781919&dl=ACM&coll=DL&CFID=540414311&CFTOKEN=26572040 |
| 6 | PLACA | 1996 | http://link.springer.com/chapter/10.1007/3-540-58855-8_23 |
| 7 | VIVA | 1996 | http:// www.informatik.uni-leipzig.de/fk/papers/viva.ps |
| 8 | DESIRE | 1997 | http://eprints.soton.ac.uk/252110/ |
| 9 | JAFMAS | 1997 | http://ieeexplore.ieee.org/xpl/login.jsp?tp=&arnumber=699236&url=http%3A%2F%2Fieeexplore.ieee.org%2Fxpls%2Fabs_all.jsp%3Farnumber%3D699236 |

**Agent-K** provides the possibility of inter-operable (or open) software agents that can communicate via KQML and which are programmed using the AOP approach.

**AGENT-0**. The initial implementation of AOP, Agent-0, is a simple language for specifying agent behavior. KQML provides a standard language for inter-agent communication.

**April** (Agent PRocess Interaction Language) is a high-level language which also offers a simple interface to other programming languages such as C. April is oriented to the implementation of agent systems; however, it is not a multiagent applications language. It does not directly offer high level features such as: planners, problem solvers and knowledge representation systems that a multiagent applications language might be expected to include. April is more a concurrent language with objects as processes.

**OASIS** (Optimal Aircraft Sequencing using Intelligent Scheduling) is a prototype air-traffic management system developed for Sydney's Kingsford Smith airport. It accurately calculates estimated landing times, determines the sequence of aircraft to land giving the least total delay, and advises air traffic controllers of appropriate control actions to achieve this sequence. It also monitors and compares actual progress of aircraft against the established sequence, and notifies the air traffic controller of significant differences and appropriate action to correct the situation. OASIS is designed to be responsive to sudden changes in environmental conditions such as meteorological conditions or runway configuration and changes in user objectives such as aircraft operational emergencies or requirements. OASIS combines artificial intelligence, software agents, and conventional software techniques.

**Agent Building Shell** provides several reusable layers of languages and services for building agent systems: description logic based knowledge management, speech-act based communication, content based information distribution, coordination modeling language, agent modeling and conflict management

**PLACA** (Planning Communicating Agents) is an agent-oriented programming language that focuses on agent planning.

**VIVA** is an agent-oriented programming language based on the theory of VIVid Agents. It follows the AOP paradigm, but adopts many concepts from SQL and Prolog.

**JAFMAS** (Java-based Agent Framework for Multi-Agent Systems) provides a generic methodology for developing speech-act based agent systems.

During the decade 2000-2009 the development of agent platforms increased, and the platforms were more diverse. Table 8 presents the most popular ones.

*Table 8. Platforms with unclear status or no longer under development (the 2000s)*

| No. | Name | LR | Website |
| --- | --- | --- | --- |
| 1 | FORR | 2000 | http://www.cs.hunter.cuny.edu/~epstein/papers/ApplyingForrToHumanMultiRobotTeams.pdf |
| 2 | JATLite | 2000 | http://www-cdr.stanford.edu/ProcessLink/papers/JATL.html |
| 3 | MINERVA | 2001 | ftp://ftp.elet.polimi.it/users/Francesco.Amigoni/pic19.pdf |
| 4 | Visual Soar | 2002 | http://web.eecs.umich.edu/~soar/sitemaker/projects/visualsoar/ |
| 5 | Concurrent MetateM | 2003 | http://cgi.csc.liv.ac.uk/~anthony/metatem.html |
| 6 | FIPA-OS | 2003 | http://fipa-os.sourceforge.net/index.htm |
| 7 | GO! | 2003 | http://dl.acm.org/citation.cfm?doid=860575.860747 |
| 8 | MAST | 2003 | http://ieeexplore.ieee.org/xpl/login.jsp?tp=&arnumber=1247718&url=http%3A%2F%2Fieeexplore.ieee.org%2Fxpls%2Fabs_all.jsp%3Farnumber%3D1247718 |
| 9 | Aglets | 2004 | http://www.research.ibm.com/topics/popups/innovate/java/html/aglets.html |
| 10 | ALIAS | 2004 | http://lia.disi.unibo.it/research/ALIAS/ <br> http://dx.doi.org/10.1023/A:1020259411066 |
| 11 | BDI4JADE | 2004 | http://www.inf.ufrgs.br/prosoft/bdi4jade/ |
| 12 | CybelePro | 2004 | http://www.i-a-i.com/cybelepro/ <br> http://www.i-a-i.com/?core/modeling-and-simulation/distributed-architectures-and-solutions <br> http://www.i-a-i.com/?core |
| 13 | IMPACT | 2005 | https://www.cs.umd.edu/projects/impact/ |
| 14 | AGLOBE, AglobeX Simulation | 2005 | https://agents.felk.cvut.cz/projects/aglobe <br> *Projects:* <br> https://agents.felk.cvut.cz/projects |
| 15 | Multiagent System Development Kit | 2005 | http://link.springer.com/chapter/10.1007%2F0-387-23152-8_9 |
| 16 | 3APL | 2006 | http://link.springer.com/chapter/10.1007/0-387-26350-0_2 |
| 17 | CAFnE | 2006 | https://researchbank.rmit.edu.au/eserv/rmit:9785/Jayatilleke.pdf (PhD thesis) |

| 18 | ZEUS | 2006 | http://sourceforge.net/projects/zeusagent/ |
|----|------|------|---------------------------------------------|
| 19 | OAA  | 2007 | http://www.ai.sri.com/~oaa/ |
| 20 | PUPS P3 | 2007 | *http://www.tumblingdice.co.uk/pupsp3/introduction.htm* |
| 21 | SOCS-SI | 2007 | http://lia.disi.unibo.it/research/projects/socs/ <br> http://dx.doi.org/10.1080/08839510500479546 |

**FORR** is a cognitive architecture that considers the opinions of others when choosing actions.

**JATLite** (Java Agent Template, Lite) is a package of Java programs that allows users to create software agents that communicate robustly over the Internet. JATLite also provides a basic infrastructure in which agents register with an "Agent Message Router" using a name and password, connect/disconnect from the Internet, send and receive messages, transfer files with FTP, and generally exchange information with other agents on the various computers where they are running.

**MINERVA** is an architecture supporting the logic-based and BDI paradigms.

**Visual Soar** is a development environment written in Java to aid in the creation of agents for use in Soar.

**Concurrent MetateM** is an agent-oriented programming language that is based on the theory of temporal logic and is influenced by methods of formal software development.

**FIPA-OS** is a component-based toolkit enabling rapid development of FIPA compliant agents. It supports the majority of the FIPA Experimental specifications.

**GO!** is a logic-based programming language.

**MAST**. Programmed in Java and built on top of the JADE agent platform, the tool is designed mainly for the simulations of material handling systems.

**Aglets** are programmed to leave their owner's computer on request, roam the Internet, and visit other computers, where they perform specific tasks such as collecting data and interacting with other agents.

**ALIAS** offers coordination of hypothetical reasoning among agents, based on abductive logic programming.

**BDI4JADE** provides an implementation of the BDI architecture, which is used as a layer on top of JADE. A key goal of this platform is to provide an environment to implement enterprise applications, which means full integration of standard technologies. The platform leverages the JADE infrastructure, such as distribution and message exchange. Recently, it was extended to provide capability relationships, which promote agent modularity.

**CybelePro** is a commercial agent infrastructure used by the government, industry and academia for applications such as robotics, planning and scheduling, data mining, modeling and simulation, and control of air and ground transportation systems, communication networks and cross-enterprise systems.

**IMPACT** is a project that aims to develop both a theory and a software implementation that facilitates the creation, deployment, interaction, and collaborative aspects of software agents in a heterogeneous, distributed environment. It provides a set of servers (yellow pages, thesaurus, registration, type and interface) that facilitate agent interoperability in an application independent manner. It also provides an Agent Development Environment for creating, testing, and deploying agents. It uses a logic-based custom syntax for agent definition.

**AGLOBE** is an agent platform designed for testing experimental scenarios featuring agents position and communication inaccessibility, but it can be also used without these extended functions. The platform focuses on the modeling and development of decentralized agent systems. It implements an efficient message transport layer, agent life-cycle management and a high level of scalability.

**Multi Agent System Development Kit** implements the Gaia methodology and supports the whole life cycle of agent system development.

**3APL** language is motivated by cognitive agent architectures and provides programming constructs to implement individual agents in terms of beliefs, goals, plans, actions, and practical reasoning rules.

**CAFnE** enables domain experts to work with a graphical representation of the system. The model of the system, updated by domain experts, is then transformed into executable code using a transformation function.

**ZEUS** provides a graphical environment to build distributed agent systems. A rule engine, planner and visualization tools are included. It contains some extensions for the DAML semantic web project and Web Services integration features.

**OAA** is a framework for integrating a community of heterogeneous software agents in a distributed environment. The architecture contains a central blackboard server holding a list of tasks, while a group of agent machines executes these tasks based on their specific capabilities.

**PUPS P3** is a portable cluster computing environment that facilitates the development of complex multi-process and multi-host computations by emulating colonies of homeostatically regulated organisms. Its central philosophy is that each process is a digital artificial life form, responsible for maintaining its own environment.

**SOCS-SI** is a runtime monitoring of social interactions among agents, which verifies that agents are complying with protocols specified in the SCIFF language.

Continuing with the subsequent decade, Table 9 shows the platforms that seem to be inactive. For some of them we provide links to projects implemented using those platforms.

*Table 9. Recent platforms that seem to be inactive (the 2010s)*

| No. | Name | LR | Website |
|---|---|---|---|
| 1 | Golog, GTGolog | 2010 | http://www.kr.tuwien.ac.at/staff/lukasiew/ki06a.pdf |
| 2 | AgentBuilder | 2011 | http://www.agentbuilder.com/ |
| 3 | AgentScape | 2011 | http://www.agentscape.org/ |
| 4 | eXAT | 2011 | http://www.erlang.org/euc/05/Santoro.pdf<br>https://github.com/gleber/exat |
| 5 | Axum | 2012 | http://social.technet.microsoft.com/wiki/contents/articles/6615.microsoft-axum-formerly-maestro.aspx |
| 6 | BRAHMS | 2012 | https://www.nasa.gov/centers/ames/research/lifeonearth/lifeonearth-brahms.html |
| 7 | CLAIM, S-CLAIM | 2012 | http://www.sciencedirect.com/science/article/pii/S1877050912003651<br>http://aimas.cs.pub.ro/people/andrei.olaru/art/2012-ANT-SCLAIMAnAgentBasedProgrammingLanguageForAmISmartRoomCaseStudy-presentation.pdf |
| 8 | MAPS | 2012 | http://maps.deis.unical.it/ |
| 9 | REC | 2012 | http://www.inf.unibz.it/~montali/tools.html#jREC<br>http://ijcai.org/papers09/Papers/IJCAI09-026.pdf |
| 10 | agentTool | 2013 | http://agenttool.cs.ksu.edu/ |
| 11 | Cougaar, ActiveEdge | 2013 | http://sourceforge.net/projects/cougaar<br>Projects:<br>http://www.cougaarsoftware.com/supported-industries/ |
| 12 | SWARM | 2013 | http://www.swarm.org http://savannah.nongnu.org/projects/swarm |
| 13 | 2APL | 2014 | http://apapl.sourceforge.net/ |

| 14 | **Event-B, Rodin** | 2014 | www.event-b.org<br>Formalizing multiagent systems example:<br>http://wiki.event-b.org/images/ZGraja-RodinWorkshop2014.pdf<br>Projects:<br>http://wiki.event-b.org/index.php/Industrial_Projects |
|---|---|---|---|
| 15 | **KGP** | 2014 | http://dx.doi.org/10.1613/jair.2596<br>*Projects:*<br>http://lia.disi.unibo.it/research/projects/socs/ |
| 16 | **RETSINA** | 2014 | http://www.cs.cmu.edu/~softagents/retsina_agent_arch.html<br>http://www.cs.cmu.edu/~softagents/software.html<br>http://www.cs.cmu.edu/~./blangley/commusersguide/Overview.html<br>Projects/Applications:<br>http://www.cs.cmu.edu/~softagents/retsina_agent_arch.html<br>http://www.cs.cmu.edu/~softagents/software.html |
| 17 | **Eve** | 2015 | https://eve.almende.com/ |
| 18 | **PRESAGE2** | 2015 | https://github.com/Presage/Presage2 |
| 19 | **Agent Factory** | 2016 | http://sourceforge.net/projects/agentfactory/<br>http://www.agentfactory.com/index.php/Main_Page<br>http://astralanguage.com/wordpress<br>*Projects:*<br>http://www.agentfactory.com/index.php/Projects |
| 20 | **Siebog** | 2016 | https://github.com/gcvt/siebog<br>https://github.com/milanvidakovic/siebog |

**GTGolog** combines explicit agent programming in Golog with multiagent planning in stochastic games.

**AgentBuilder** is an integrated tool suite for constructing intelligent software agents. It has two major components: the "Toolkit" and the "Run-Time System". The "Toolkit" includes tools for managing the agent-based software development process, analyzing the domain of agent operations, designing and developing networks of communicating agents, defining behaviors of individual agents, and debugging and testing agent software. The "Run-Time System" includes an agent engine that provides an environment for the execution of agent software.

**AgentScape** is a middleware layer that supports large-scale agent systems. It supports multiple code bases and operating systems, and interoperability with other agent platforms.

**eXAT** (Erlang eXperimental Agent Tool) provides an "all-in-one framework" allowing to design, with a single tool, agent intelligence, agent behavior and agent communication. This is made possible by means of a set of modules strongly tied to one another: an Erlang-based expert system engine, an execution environment for agent behaviors based on object-oriented finite-state machines, and a module able to handle FIPA-ACL messages.

**Axum** is a domain specific concurrent programming language, based on the Actor model. It is an object-oriented language based on the .NET Common Language Runtime using a C-like syntax which, as a domain-specific language, is intended for developing portions of a software application that is well-suited to concurrency. An agent (or an actor) is an isolated entity that executes in parallel with other agents.

**BRAHMS**, created by NASA, is a set of software tools to develop and simulate multiagent models of human and machine behavior. It was originally developed to analyze or design human organizations and work processes. Brahms is a multiagent, rule-based, activity programming language. It has similarities to the BDI architecture and other agent-oriented languages, but is based on a theory of work practice and situated cognition. It also integrates the subsumption architecture.

**CLAIM, S-CLAIM** are platforms built in JADE, providing a Lisp-like language.

**MAPS** (Mobile Agent Platform for Sun SPOTs) is a Java-based framework for wireless sensor network (WSNs) based on Sun SPOT technology which enables agent-oriented programming of WSN applications. The MAPS architecture is based on components which interact through events. Each component offers a minimal set of services to mobile agents which are modeled as multi-plane state machines driven by ECA rules. In particular, the offered services include message transmission, agent creation, agent cloning, agent migration, timer handling, and easy access to the sensor node resources.

**REC** is a runtime monitoring of compliance of agents to commitment. It is based on reactive event calculus.

**agentTool** is a Java-based graphical development environment to help users analyze, design and implement agent systems. It is designed to support the Organization-based Multiagent Systems Engineering (O-MaSE) methodology.

**Cougaar, ActiveEdge** (Cognitive Agent Architecture) is a Java-based architecture for the construction of large-scale distributed agent-based applications. It is a product of two consecutive, multi-year DARPA research programs into large-scale agent systems. ActiveEdge is an intelligent decision support platform. It uses a distributed intelligent agent architecture based on the human cognitive model of reasoning and planning, which captures the way humans observe, reason, plan and act.

**SWARM** is a kernel and library for the multiagent simulation of complex systems. The basic architecture of Swarm is a collection of concurrently interacting agents.

**2APL** (A Practical Agent Programming Language) is a modular BDI-based programming language that supports the development of agent systems. 2APL provides a set of programming constructs allowing direct implementation of concepts such as beliefs, declarative goals, actions, plans, events, and reasoning rules. The reasoning rules allow run-time selection and generation of plans based on declarative goals, received events and messages, and failed plans. It can be used to implement agent systems consisting of software agents with reactive as well as proactive behaviors. The platform can be used in two modes: in the stand-alone mode or in distributed mode using the JADE platform.

**Event-B, Rodin** is a formal method for system-level modeling and analysis. Key features are the use of set theory as a modeling notation, the use of refinement to represent systems at different abstraction levels and the use of mathematical proof to verify consistency between refinement levels. Rodin is the name of the tool platform for Event-B. It allows formal Event-B models to be created with an editor.

**KGP** is an agent modeling framework entirely based on computational logic.

**RETSINA** (Reusable Environment for Task-Structured Intelligent Networked Agents) is an agent architecture developed at Software Agents Lab at Carnegie Mellon University's Robotics Institute.

**Eve** is a multipurpose, web-based agent platform. It aims to be an open and dynamic environment where agents can live and act anywhere: in the cloud, on smartphones, on desktops, in browsers, robots, home automation devices, and others. The agents communicate with each other using simple, existing protocols (JSON-RPC) over existing transport layers (HTTP, XMPP), offering a language- and platform-agnostic solution.

**Presage2** is a simulation platform for agent-based simulation. It enables designers to investigate the effect of agent design, network properties and the physical environment on individual agent behavior and long-term collective global performance.

**Agent Factory** is a open-source, FIPA-based collection of tools, platforms and languages that support the development and deployment of agent systems. ASTRA is an agent programming language that is built on and integrated with Java. This is an implementation of AgentSpeak(TR), a logic-based agent programming language that combines AgentSpeak(L) with teleo-reactive functions.

**Siebog** is an enterprise-scale multiagent middleware consisting of the following main modules: an extensible Java EE-based Agent Framework operating on top of computer clusters, offering automatic agent load-balancing, state-replication, and fault-tolerance; Radigost, a web-based multiagent platform, built using JavaScript and HTML5-related standards. Radigost agents are executed inside web browsers and can be used in a wide variety of hardware and software platforms, including personal computers, smartphones and tablets, Smart TVs, etc.; a Jason interpreter, a port of the popular Jason interpreter to Java EE; and a distributed non-axiomatic reasoning system, an advanced reasoning system based on the non-axiomatic logic formalism.

## 6. Conclusions

Agent systems provide a bottom up approach for analyzing complex systems. Agents can be the means for modeling and simulating diverse phenomena which are difficult to model and understand using traditional, analytical methods. Because of this research interest, a large number of platforms specifically designed to ease the programming of agent-based applications have been proposed. Although the main applications of agent systems belong to the computer science field, and often related to artificial intelligence, there are increasingly many uses found in areas such as life sciences, ecological sciences and social sciences.

In this work, we presented a detailed review of the available platforms, and also of those which are no longer actively being developed, but can be given merit from a historical perspective. Our main goal was to help researchers in assessing their characteristics in order to choose those which best suit their scientific necessities. In this respect, we tried to perform an extensive inventory of the platforms and present their key features. We tried our best to include the most significant ones, but it is certainly possible that there are other popular platforms missing. It is also possible that some platforms presented as legacy are still active.

Also, the specific details of various agent platforms can give the interested reader a broad perspective of the concepts and techniques used in agent systems and perhaps encourage new theoretical and practical developments in this field.